\documentclass[a4paper,11pt]{article}

\usepackage{jheppub} 
\usepackage{physics}

\usepackage[T1]{fontenc} 

\title{\boldmath $1/N$ correction in holographic Wilson loop  
from quantum gravity}

\author{Koji Hashimoto,}
\author{Wataru Sasaki,}
\author{Takayuki Sumimoto}

\affiliation{Department of Physics, Osaka University,\\1-1 Machikaneyama, Toyonaka, Osaka 560-0043, Japan}

\emailAdd{koji@phys.sci.osaka-u.ac.jp}
\emailAdd{wataru.sasaki@het.phys.sci.osaka-u.ac.jp}
\emailAdd{t\_sumimoto@het.phys.sci.osaka-u.ac.jp}

\abstract{
We study $1/N$ corrections to a Wilson loop in holographic duality. Extending the AdS/CFT correspondence beyond
the large $N$ limit is an important but a subtle issue, as it needs quantum gravity corrections in the gravity side.
To find a physical property of the quantum corrected geometry of near-horizon black 0-branes
previously obtained by Hyakutake, we evaluate a Euclidean string worldsheet hanging down in the geometry,
which corresponds to a rectangular Wilson loop in the $SU(N)$ quantum mechanics with 16 supercharges
at a finite temperature with finite $N$. We find that the potential energy defined by the Wilson loop
increases due to the $1/N$ correction, therefore the quantum gravity correction weakens the gravitational attraction.
}

\begin{document} 
\maketitle
\flushbottom

\section{Introduction}
\label{sec:intro}

Whether the AdS/CFT correspondence \cite{Maldacena:1997re,Gubser:1998bc,Witten:1998qj}
works beyond the large $N$ limit is one of the central problems in holography and
quantum gravity. This is simply because the question leads to the consistency of having the holographic principle 
as a definition of quantum gravity. In spite of the importance, the difficulty in calculating the
quantum gravity corrections to the AdS/CFT observables prevented us from the explicit check of
the validity of the AdS/CFT beyond the large $N$ limit.

A nontrivial check was made in Ref.~\cite{Hanada:2013rga,Hanada:2016zxj} at which 
the energy of the near horizon geometry of black 0-branes was shown to coincide with that of the dual $SU(N)$
supersymmetric quantum mechanics at finite temperature, to the first nontrivial correction 
order in the $1/N^2$ expansion.
The analysis on the gravity side was based on the quantum-corrected near horizon geometry obtained by Hyakutake
\cite{Hyakutake:2013vwa}, therefore this nontrivial result shows that the geometry is a good test ground 
for the AdS/CFT correspondence beyond the large $N$ limit.\footnote{To obtain nontrivial $1/N$ terms in
the gravity side, one needs to break at least a part of supersymmetries, as $AdS_5\times S^5$ is expected
to have no correction due to the largest number of supersymmetries. See for example Ref.~\cite{Ardehali:2013gra}
for the check of the spectra at the $1/N^2$ level for an orbifolded $S^5$. See also Ref.~\cite{Honda:2019cio}
for recent discussions on the $1/N$ corrections to the $AdS_5$ blackhole entropy.}

The finite $N$ AdS/CFT offers interesting issues, among which here we mention the phase transition.
The Hawking-Page transition \cite{Hawking:1982dh}
in AdS/CFT is known \cite{Witten:1998zw} to correspond to confinement/deconfinement transition
which is a first order transition at the large $N$ limit. On the other hand, theories such as QCD at $N=3$
has a cross-over phase transition, thus the finite $N$ drastically changes the thermal behavior.
Even if one includes the $1/N^2$ correction to the gravity metric, intuitively the Hawking-Page transition is
expected to persist, and the phase transition is still at the first order, which causes an issue about
how the finite $N$ effect should be seen in the gravity side. Interestingly, a gravity geometry
inversely obtained to reproduce QCD lattice data \cite{Hashimoto:2018bnb} using a deep learning technique
\cite{Hashimoto:2018ftp} has a weak repulsive wall just outside the black hole horizon, suggesting 
the coexistence of confinement and deconfinement phases in gravity.\footnote{The repulsive wall structure
was also found in a probe D0-brane potential in the quantum corrected geometry \cite{Hyakutake:2013vwa}.} 

To find any local structure of the quantum-corrected black hole geometry, we need some probe. In this paper 
we evaluate a holographic Wilson loop of the $SU(N)$ quantum mechanics with 16 supercharges 
at a finite temperature. The gravity dual is a Euclidean string worldsheet ending on the contour of the Wilson loop
\cite{Maldacena:1998im}. Generally the rectangular Wilson loop is related to a potential energy of 
a quark and an antiquark pair, and here we consider the W-bosons, by separating two of the D0-branes
from the remaining $N$ D0-branes which compose the black hole\footnote{
Since we treat D0-branes, there is only one direction in their worldvolume: time. To form a rectangular
Wilson loop, the other direction is necessary, and here we take it to be a direction in $S^ 8$ which is spanned by
the scalar field of the quantum mechanics. Therefore
the two separated D0-branes are located in different positions. This kind of situation was treated originally in 
Ref.~\cite{Maldacena:1998im}.}.  
The quantum gravity corrections due to the geometry
provides the $1/N^2$ corrections to the potential energy. We will find that the resultant contribution to
the potential energy is positive, meaning that the gravitational attraction is weakened due to the
quantum gravity corrections. This appears to be consistent with the general argument above to make the
phase transition weaker.

In addition to the correction due to the bulk quantum effect which we treat in this paper, 
as for the Wilson loop there exists another correction of the same order ${\cal O}(1/N^2)$, that is a
higher genus correction of the Euclidean string worldsheet for the probe. 
This effect has been explicitly calculated in \cite{Forini:2017whz} for the case of a circular Wilson loop\footnote{ 
See also Refs.~\cite{Forini:2015bgo, Faraggi:2016ekd} for the history of the matching in the AdS/CFT. 
For circular Wilson loops with a totally symmetric/antisymmetric representation of $SU(N)$ whose dual is not a string
but a D3-brane/D5-brane, see Refs.~\cite{Buchbinder:2014nia,Faraggi:2014tna}.}. 
This effect, although at the same order, is difficult to calculate in our case and so we 
concentrate on the bulk quantum gravity correction.

The organization of this paper is as follows. First in Sec.~\ref{sec2}, we review the quantum-corrected gravity geometry of the black 0-branes in type IIA string theory given by Hyakutake \cite{Hyakutake:2013vwa}, 
and see that the parameters,  $N$ and the temperature, needs to satisfy certain inequalities so that the
quantum corrections are dominant against $\alpha'$ corrections. Furthermore, we study the reality condition
of the spacetime metric, and the spacetime region where the analysis is valid. 
In Sec.~\ref{sec3}, we evaluate the Wilson loop and its $1/N$ correction due to the quantum correction
of the spacetime geometry in the gravity side. The obtained potential energy shows that the 
attractive force of the gravity is weakened by the $1/N$ correction. Sec.~\ref{sec4} is for our summary and
discussions, and Appendix A is a review of the details of the quantum corrected geometry \cite{Hyakutake:2013vwa} which we use in this paper.


\section{Spacetime conditions}
\label{sec2}

In this section, we briefly review the result of \cite{Hyakutake:2013vwa} for introducing
the target space of our interest,
and analyze necessary conditions for our worldsheet analyses in the later sections to be valid.

\subsection{Review: Quantum correction of the near horizon black 0-brane geometry}
\label{sec2-1}

Here we introduce the quantum-corrected near horizon geometry of the black $N$ 0-branes obtained in 
Ref.~\cite{Hyakutake:2013vwa}. Let us describe first the geometry without the quantum correction,
that is a near-horizon solution of the type IIA supergravity \cite{Itzhaki:1998dd}:
    \begin{align}\label{classicalmetric}
      ds^{2} &= l_{s}^{2}\left(-H^{-1/2}Fdt^{2}+H^{1/2}F^{-1}dU^{2}+H^{1/2}, 
      U^{2}d\Omega_{8}^{2} \right),  \quad
      e^{\phi}=l_{s}^{-3}H^{3/4}, \\
      H &\equiv \frac{(2\pi)^{4}15\pi\lambda}{U^{7}}, \quad
      F\equiv 1-\frac{U_0^7}{U^7}.
    \end{align}
Here 
$\phi$ is the dilaton field, while the Ramond-Ramond 1-form is omitted. 
$\lambda$ is the 't Hooft coupling of the dual gauge theory, the $SU(N)$ quantum mechanics with 16 supercharges
The radial coordinate is $U\equiv r/l_s^2$ in which we rescaled the original supergravity coordinate $r$ by the string length $l_s$, thus having the dimension of energy. $U_0$ is a parameter determining the event horizon of the geometry and resultantly the temperature, 
    \begin{align}
      \tilde{U}_0 &\equiv \frac{U_0}{\lambda^{1/3}}
      = \qty({\frac{7}{16\pi^3 \sqrt{15\pi}}})^{-2/5}\tilde{T}^{2/5}.
    \end{align}
The dimensionless coordinate $\tilde{U}_0$ and the dimensionless temperature
$\tilde{T}\equiv T/\lambda^{1/3}$ were introduced for convenience.

According to Ref.~\cite{Hyakutake:2013vwa}, in order to obtain the quantum-corrected geometry,
one makes use of the fact that the black 0-brane solution \eqref{classicalmetric}
is a dimensionally reduced
M-wave solution \cite{Cremmer:1978km} in 11-dimensional supergravity. At the 11 dimensions,
some higher derivative corrections to the Einstein-Hilbert action are known,
thus derive the 1-loop quantum corrections to the type IIA supergravity in string theory.
The geometry solving the corrected 11-dimensional supergravity provides the quantum corrected
black 0-brane solution at the dimensional reduction. The obtained geometry is \cite{Hyakutake:2013vwa}
    \begin{align}
      \label{hyakutakemetric}
      ds^{2} &= l_{s}^{2}\left(-H_{1}^{-1}H_{2}^{1/2}F_{1}dt^{2}+H_{2}^{1/2}F_{1}^{-1}U_{0}^{2}dx^{2}+H_{2}^{1/2}U_{0}^{2}x^{2}d\Omega_{8}^{2}\right) ,\\
      \label{hyakutakedilaton}
      e^{\phi}&=l_{s}^{-3}H_{2}^{3/4}.
    \end{align}
Here $x\equiv U/U_0$ is a dimensionless radial coordinate, and the functions 
$H_i,F_1$ are defined in Appendix A. In the limit $N\to\infty$, those functions reduce to
the original uncorrected geometry \eqref{classicalmetric}, consistently. Note that 
the location of the 
horizon of the geometry, originally at $x=1$, is shifted to some $x\neq 1$ in this coordinate system.

\subsection{Constraints on the temperature and $N$}
\label{sec2-2}

The higher derivative corrections of the gravity consist of quantum corrections and stringy corrections. 
In Ref.~\cite{Hyakutake:2013vwa}, a condition for finding that the quantum corrections are dominant
against the stringy corrections was calculated, 
    \begin{align}
      \tilde{T}<s^{5/9},\qq{} \sqrt{\frac{c_2}{c_1s}}\tilde{T}^{-3/2} &< N <\sqrt{c_1s}\, \tilde{T}^{-21/10}. 
      \label{N-Tregion1}
    \end{align}
These inequalities are derived by the comparison of the various terms in the entropy of the black hole.\footnote{In
Ref.~\cite{Hyakutake:2013vwa} the comparison was claimed to be made for the internal energy $E$, 
assuming a relation $E=TS$.}
The parameter $s \,(\ll 1)$ introduced here is a small constant  to quantify the hierarchy among 
the magnitudes of the correction terms. 
In this paper we use 
$s=0.1$ below, together with $c_1=0.779$ \cite{Hyakutake:2013vwa}
and $c_2=0.00459$ \cite{Hanada:2013rga}.
We note that the heat capacity of the geometries \eqref{hyakutakemetric} and \eqref{classicalmetric}
is positive when \eqref{N-Tregion1} is satisfied, so these spacetimes are thermodynamically stable
\cite{Hyakutake:2013vwa}.

In addition to the inequalities \eqref{N-Tregion1} found in Ref.~\cite{Hyakutake:2013vwa}, we find 
a new inequality to be satisfied for a consistent geometry. The location of the horizon
$x=x_h$ is determined by a solution of the equation $F_1(x_h)=0$, while the other metric factors
$H_i(x)$ could vanish in the region $x>x_h$. This leads to an inconsistency, because $H_i(x)=0$
means that the perturbative correction is comparable to the original uncorrected metric.
Furthermore, since the metric \eqref{hyakutakemetric} includes a square root of $H_i(x)$, 
the metric becomes imaginary. Therefore we have to require 
    \begin{align}
      H_i(x)>0 \quad \text{for} \quad  x > x_h \label{N-Tregion2}
    \end{align}
for our geometry and the corrections to be consistent. This inequality is rephrased 
in terms of $N$ and $\tilde{T}$. 
A straightforward numerical evaluation shows that a condition 
\begin{align}
H_2(x_h)>0
\label{H2pos}
\end{align}
ensures the constraints for the metric functions \eqref{N-Tregion2} (see Appendix A). So we choose this inequality \eqref{H2pos} as a necessary condition for $(\tilde{T},N)$ to be satisfied. 

\begin{figure}[t]
      \centering
      \includegraphics[width=.6\linewidth]{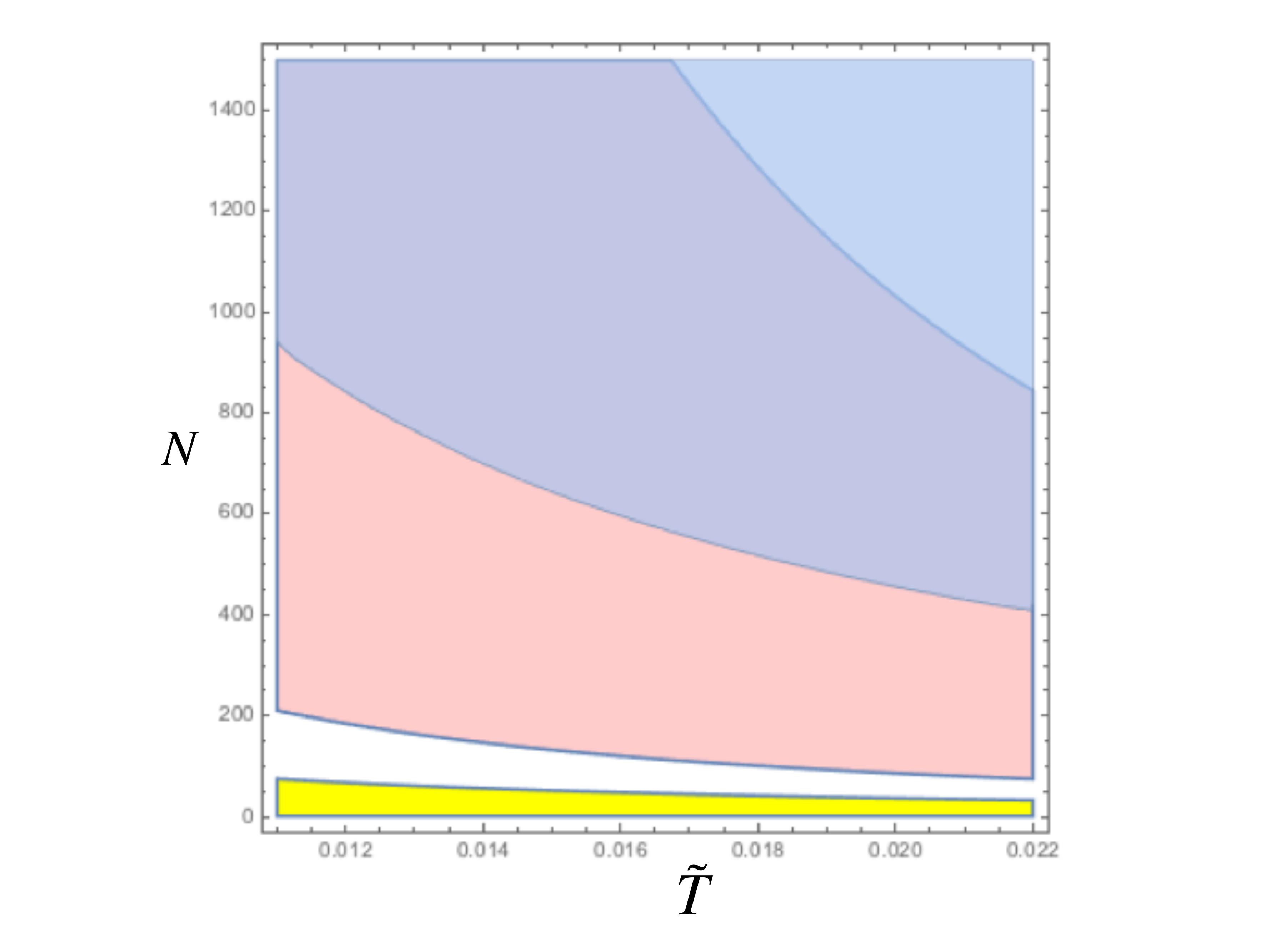}
      \caption{
      Allowed regions in $(\tilde{T},N)$ plot.
      Blue region shows that for $H_2(x_h)>0$. Red region shows \eqref{N-Tregion1} with $s=0.1$. 
      We also show the region (Yellow) where the heat capacity is negative and thus the spacetime is thermodynamically unstable.
      }
      \label{fig_NTregion}
    \end{figure}

In Fig.\ref{fig_NTregion}, we plot all the inequalities including \eqref{N-Tregion1}. We have to choose a set of values $(\tilde{T},N)$ from the allowed region shown in Fig.\ref{fig_NTregion}.

\subsection{Valid spacetime region}
\label{sec2-3}

Up to now we have looked at some necessary conditions for the quantum gravity corrections to be
a dominant correction and for the spacetime to make sense.
In this subsection, we obtain conditions about valid spacetime regions, by assuming that string perturbation makes sense.

As is found in the supergravity metric \eqref{classicalmetric}, black $p$-brane 
supergravity solutions except for $p=3$ are not scale invariant, and resultantly 
there appears spacetime regions where quantum gravity corrections or stringy corrections
grow to overwhelm the consistency of string perturbation theory. 
Perturbation consistency requires
    \begin{align}
  		l_s^2/\rho^{2}&\ll 1,  \label{sugra_conditon1} \\
  		g_se^{\phi}&\ll 1. \label{sugra_conditon2}
  	\end{align}
The first inequality \eqref{sugra_conditon1} means that the typical gravitational curvature scale $\rho$ is much larger than 
the string scale so that stringy corrections are perturbative, and the second inequality \eqref{sugra_conditon2} means
that the string coupling
is small enough such that string perturbation theory makes sense.

In our case, as $\rho$ we may take the curvature scale of $S^8$, so using the solution \eqref{hyakutakemetric} we rephrase \eqref{sugra_conditon1} and \eqref{sugra_conditon2} as
    \begin{align}
  		\qty[H_{2}(x){}^{1/2} U_{0}^2 x^{2}]^{-1} & \ll 1,  \label{x_ucondition} \\
      \frac{(2\pi)^{2}\lambda}{N}H_{2}(x)^{3/4}& \ll 1. \label{xdowncondition}
  	\end{align}
These provides a region in the spacetime at which the string perturbation makes sense.
Because the left hand side of \eqref{x_ucondition} has a positive power in $x$ and
that of \eqref{xdowncondition} has a negative one, these inequalities provide 
the upper bound and the lower bound of $x$, respectively.

\begin{figure}[tp]
    \centering
      \includegraphics[width=.49\linewidth]{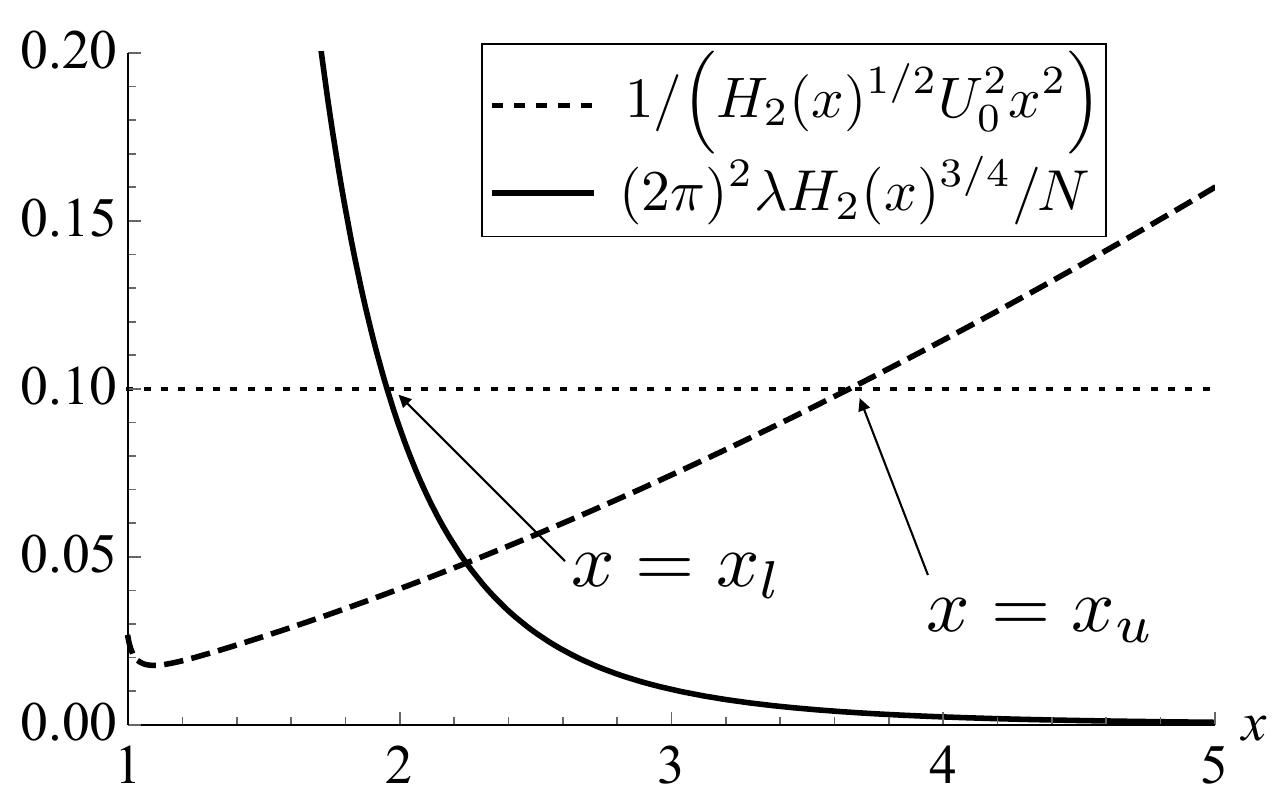}
      \hspace{5mm}
      \includegraphics[width=.45\linewidth]{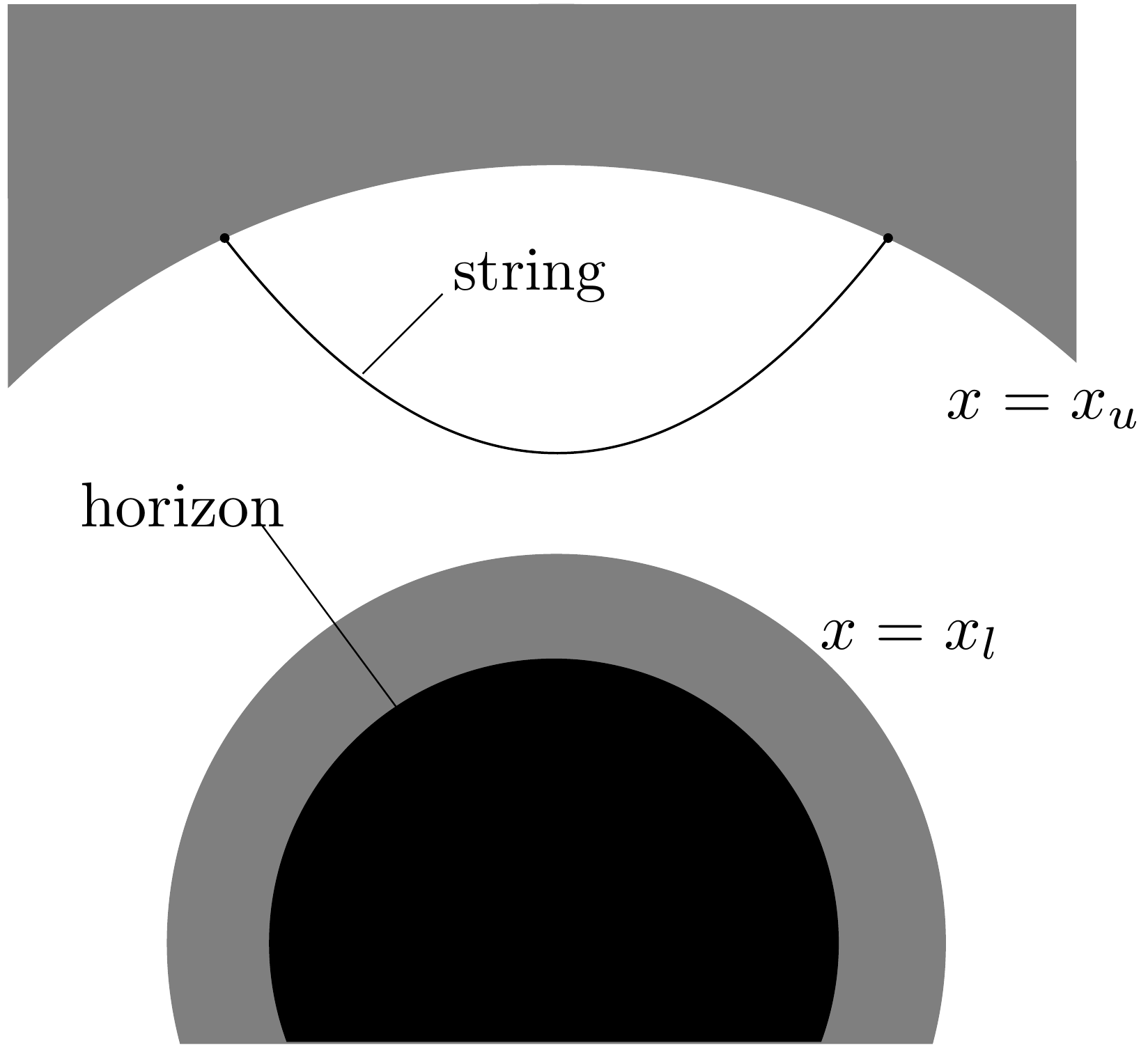}
      \caption{Left: Plot of the left hand sides of \eqref{x_ucondition} and \eqref{xdowncondition}, for $N  = 456$ and $\tilde{T} =0.02$. A larger $N$ results in a wider region in $x$.
      Right: Valid spacetime region (White) and our probe string configuration. The shaded regions are excluded for use as the
      string perturbation breaks down.
      }
    \label{fig_region}
    \end{figure}

To evaluate the spacetime region concretely for our later purpose, we assume 
a threshold value $r \,(<1)$ for the left hand sides of the inequalities and
define the upper bound $x_u$ and the lower bound $x_l$ for the radial coordinate $x$,\footnote{These two bounds can
be parameterized by different values, but here for simplicity we parameterize them with the single parameter $r$.}   
    \begin{align}
  		H_{2}(x_u){}^{1/2} U_{0}^2 x_u^{2} & = \frac{1}{r},  \label{x_u} \\
      \frac{(2\pi)^{2}\lambda}{N}H_{2}(x_l)^{3/4} & = r. \label{x_l}
  	\end{align}
Solutions of these equations for a given small $r$, we obtain a spacetime region 
\begin{align}
x_u>x>x_l
\label{xul}
\end{align}
at which the string perturbation expansion makes sense.\footnote{For the uncorrected geometry
\eqref{classicalmetric}, $H_2(x)$ in these conditions is simply replaced by $H(x)$.}
Fig.~\ref{fig_region} Left explains the inequalities for our choice of 
$N=456$ and $\tilde{T} =0.02$ with $r=0.1$.

In summary, the quantum-corrected geometry \eqref{hyakutakemetric} can be used for
$(\tilde{T},N)$ satisfying \eqref{N-Tregion1} and \eqref{H2pos}, in the region \eqref{xul}.
In the next section, we calculate the holographic Wilson loop by putting a probe string
in the valid region of the spacetime given by a valid $(\tilde{T},N)$.

\section{$1/N$ corrections to Wilson loop}
\label{sec3}

In this section, we calculate a holographic Wilson loop of the supersymmetric quantum mechanics
at a finite temperature, using a string worldsheet in the quantum-corrected geometry \eqref{hyakutakemetric}.

\subsection{Set-up for the Wilson loop}

We follow the original proposal \cite{Maldacena:1998im,Rey:1998ik} for the holographic 
calculation of the Wilson loop. 
We have to make sure that the string worldsheet in the curved geometry exists only in the region where
the calculation is valid, \eqref{xul}. For this purpose, we put two probe D-branes at $x=x_u$, and consider
a fundamental string hanging down from there. See Fig.~\ref{fig_region} Right. 
These D-branes can be D0-branes. In that case, the Wilson loop we consider is
in a background of a nontrivial scalar expectation value: 
\begin{align}
\Phi^I = \left(
\begin{array}{ccccc}
0 & & & 0 & 0\\
& \cdots & & 0 & 0\\
& & 0 & 0 & 0\\
0 & 0 & 0 & \phi_1^I & 0 \\
0 & 0 & 0 & 0 & \phi_2^I 
\end{array}
\right).
\end{align}
The $(N+2)\times(N+2)$ matrix $\Phi^I$ is separated into $SU(N)$ sector and two $U(1)$ sectors, and the vacuum expectation values $\phi_1^I$ and $\phi_2^I$ $(I=1,2,\cdots,9)$ provide the location of the two probe D0-branes.\footnote{
There should be a finite-temperature potential
between the probe D0-branes and the $N$ D0-branes, therefore the probe D0-branes are not stable. 
This problem would be absent for the case of other kinds of D-branes with different dimensionality.
For example, one can additionally place probe D4-branes which form a bound state with the probe D0-branes,
so that the location of the D0-branes in the target space can be fixed, like a flavor D-brane \cite{Karch:2002sh}.
}
We assume $(\phi_1^I)^2 = (\phi_2^I)^2$, and the angle between the two vectors $\phi_1$ and $\phi_2$ is denoted as 
$\phi$.

The Wilson loop operator relevant to this static string worldsheet is \cite{Maldacena:1998im}
    \begin{align} \label{Wilsonloop1}
      W(C) = \frac{1}{N} \tr P \exp \qty[ i \oint_C ds 
      \left(\frac{dt}{ds} A_0(t)+ \theta^I(s) \Phi^I (t) \sqrt{|dt/ds|^2} \right)]
    \end{align}
where the contour $C$ has a rectangular shape in $x^0$ and $\theta$ spacetime, of the size $\tau \times \phi$.
Here, 
$\theta$ is a coordinate for the
equator of the spherical target space $S^8$ spanned by the angular part of the nine scalar fields $\Phi^I$.
So, when $dt/ds\neq 0$, $\theta^I(s)$ takes either $\phi_1^I/|\phi_1|$ or $\phi_2^I/|\phi_2|$.
This generalized Wilson loop corresponding to a fundamental string hanging down from the two separate probe D-branes
was first considered in Ref.~\cite{Maldacena:1998im}. 
We take a long rectangle limit $\tau \to \infty$ so that the Wilson loop has the meaning of the potential energy
between two ``W-bosons.''

The location of the two D0-branes is given by the vacuum expectation values $\phi_1^I$ and $\phi_2^I$. To place these 
at $x=x_u$ in the holographic spacetime, we need a physical relation between the scalar field and the spacetime coordinate.
The definition of $x=x_u$ is given in \eqref{x_u}. This equation means that the curvature of the spacetime geometry
is not large compared to stringy scale. More precisely, the amplitude of the worldsheet instanton wrapping an 
$S^2$ of the $S^8$ is 
    \begin{align}
      S_\text{inst} = \frac{1}{2\pi l^2_s} \times {\rm Vol}(S^2) = \frac{2\rho^2}{l_s^2}
    \end{align}
so the instanton correction is $\exp[-S_\text{inst}]$. This instanton correction should have 
a counterpart in the quantum mechanics, which in principle 
determines the vacuum expectation value $|\phi_1|$ and $|\phi_2|$, even with the $1/N$ 
corrections, by the relation $2/r = S_\text{inst}$.
\footnote{Note that the ordinary relation to identify $2\pi \alpha' |\phi|$ as the bulk radial coordinate
may be corrected by $1/N$ and the temperature, so in our setup to see the quantum gravity correction in
the Wilson loop we do not use the ordinary relation. To find a relation between the vacuum expectation value and
the bulk location, a physical observable is necessary, and here we propose that it could be the string worldsheet amplitudes. 
We do not know the precise analytic 
relation between our $x_u$ and $|\phi_1|$, and leave it for our future investigation.}

    \begin{figure}[t]
  		\centering
  		\includegraphics[width=0.5\linewidth]{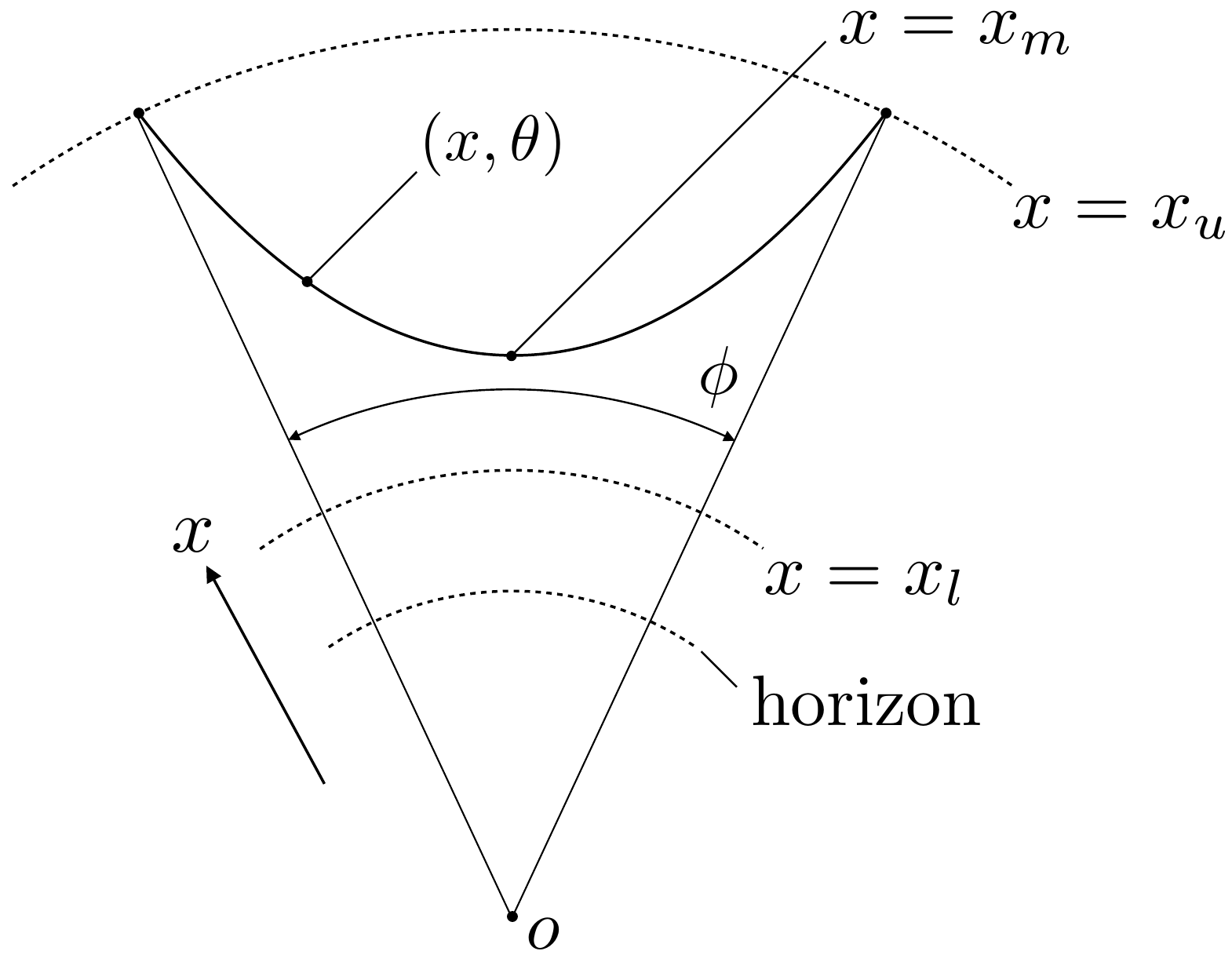}
  		\caption{The configuration of the fundamental string in the space
		spanned by the coordinates $(x,\theta)$.  The string hangs 
		down from the end points at $x=x_u$. It reaches $x=x_m$ at the bottom. 
		The bottom needs to satisfy $x_m > x_l$ so that the string lies only in the valid spacetime region.
		}
  		\label{fig_image_string}
  	\end{figure}

\subsection{Evaluation of holographic Wilson loop and $1/N$ corrections}

To calculate the string worldsheet area $S_\text{NG}$, we note that there exists a turning point $x=x_m$ which determines
the whole configuration of the string, see Fig.~\ref{fig_image_string}. The configuration of the string is given by a function
$\theta(x)$ which can be obtained by integrating the Nambu-Goto equation of motion in the background geometry.
Denoting $t_\text{E}$ as the Euclidean time, we find the worldsheet 
area of a string hanging down from the two D0-branes as
    \begin{align}
  		S_\text{NG} &= 2\times\frac{1}{2\pi l_s^2} \int_{x_m}^{x_u} dx \int^\tau_0 dt_\text{E} \sqrt{l_s^4\frac{\sqrt{H_2(x)}}{H_1(x)}F_1(x)\left(\frac{\sqrt{H_2(x)}}{F_1(x)}U_0^2+\sqrt{H_2(x)}U_0^2x^2\theta^{\prime 2}\right)} 
		\nonumber \\
      &=\frac{\tau U_0}{\pi} \int^{x_u}_{x_m}dx \sqrt{\frac{H_2(x)}{H_1(x)} \left( 1 + F_1(x)x^2\theta^{\prime 2} \right)}.
      \label{NGaction}
  	\end{align}
Here, we have defined $\theta' \equiv d\theta/dx$. The integrand does not depend on $\theta$ explicitly, so
there exists a Noether invariant
    \begin{align}
      p_{\theta}(x) &\equiv  \sqrt{ \frac{H_2(x)}{H_1(x)}} \frac{x^2 F_1(x)}{\sqrt{\theta^{\prime}(x)^{-2}+x^2 F_1(x)}} = \text{const.}
      \label{p_theta}
    \end{align}
Since at the turning point $x=x_m$ we have $\dv*{x}{\theta}=0$, this constant can be easily evaluated at $x=x_m$.
Using that, we can refine \eqref{NGaction} so that the integrand is written only as an explicit function of $x$.
The potential energy $V \equiv\lim_{\tau\to\infty }S_\text{E}/\tau$ which is related to the Wilson loop as
$\ev{W(C)} = e^{-\tau V}$ is obtained as 
    \begin{align}
  		V(x_m) = \frac{U_0}{\pi} \int^{x_u}_{x_m} dx \sqrt{\frac{H_2(x)}{H_1(x)}\left(1 + \left[\left(\frac{H_2(x)}{H_1(x)}x^2F_1(x)\right) /\left(\frac{H_{2}(x_m)}{H_{1}(x_m)}x_m^2F_{1}(x_m)\right) -1\right]^{-1}\right)}.
      \label{Vc_expression}
  	\end{align}

For this to be interpreted as an observable in the quantum mechanics, we need to calculate the relation between
$x_m$ and $\phi$. It is obtained by an integration of $\theta'$ obtained by \eqref{p_theta} along the string. We find
    \begin{equation}
  		\phi = 2 \int^{x_u}_{x_m}dx \sqrt{x^2F_1\left[  		\left(\frac{H_2(x)}{H_1(x)}x^2F_1(x)\right)/\left(\frac{H_{2}(x_m)}{H_{1}(x_m)}x_m^2F_{1}(x_m)\right)-1\right]}.
      \label{phic_expression}
  	\end{equation}
Both the potential $V$, \eqref{Vc_expression}, and the angle $\phi$, \eqref{phic_expression}, are expressed by
$x_m$, so eliminating $x_m$ from these two equations, we obtain the potential $V(\phi)$.

The purpose of our study is to compare the potential energy $V_0$ for the original large $N$ limit with the metric
\eqref{classicalmetric} and the potential energy $V$ for the metric with the $1/N$ correction, \eqref{hyakutakemetric}.
To see it, here we write the expression for the large $N$ limit:
    \begin{align}
      V_0
      &= \frac{U_{0}}{\pi} \int^{x_u}_{x_m} dx \sqrt{ \left(1+\left[\left(x^2F(x)\right)/\left(x_m^2F(x_m)\right)-1\right]^{-1} \right)}, \label{V0_expression} \\
      \phi
      &= 2 \int^{x_u}_{x_m} dx \sqrt{ x^2F(x) \left( \left(x^2F(x)\right)/\left(x_m^2F(x_m)\right) - 1 \right) }.
       \label{phi0_expression}
    \end{align}
These are obtained just by replacing $H_1(x),H_2(x)$ by $H(x)$ and $F_1(x)$ by $F(x)$.
From these expressions we can calculate the potential energy $V_0(\phi)$ for the large $N$ limit.

Let us calculate the change of the potential energy by the $1/N$ correction,
\begin{align}
\Delta V_{N,\tilde{T}}(\phi) \equiv V(\phi)-V_0(\phi).
\label{diffV}
\end{align}
Our procedures for numerically evaluating this correction are as follows. 
First, based on the argument given in Sec.~\ref{sec2-2}, we choose an appropriate set $(N,\tilde{T})$ to determine the 
spacetime geometry. Next, we calculate $x_u$ and $x_l$ by solving  \eqref{x_u} and \eqref{x_l}
to find the valid region of the spacetime, as shown in Fig.~\ref{fig_region}. In the valid region we vary $x_m$,
and numerically calculate \eqref{Vc_expression}, \eqref{phic_expression},  
\eqref{V0_expression} and \eqref{phi0_expression} for each $x_m$ comprehensively, to finally evaluate 
the quantity $\Delta V_{N,\tilde{T}}(\phi)$.

    \begin{figure}[t]
         \includegraphics[width=0.45\textwidth]{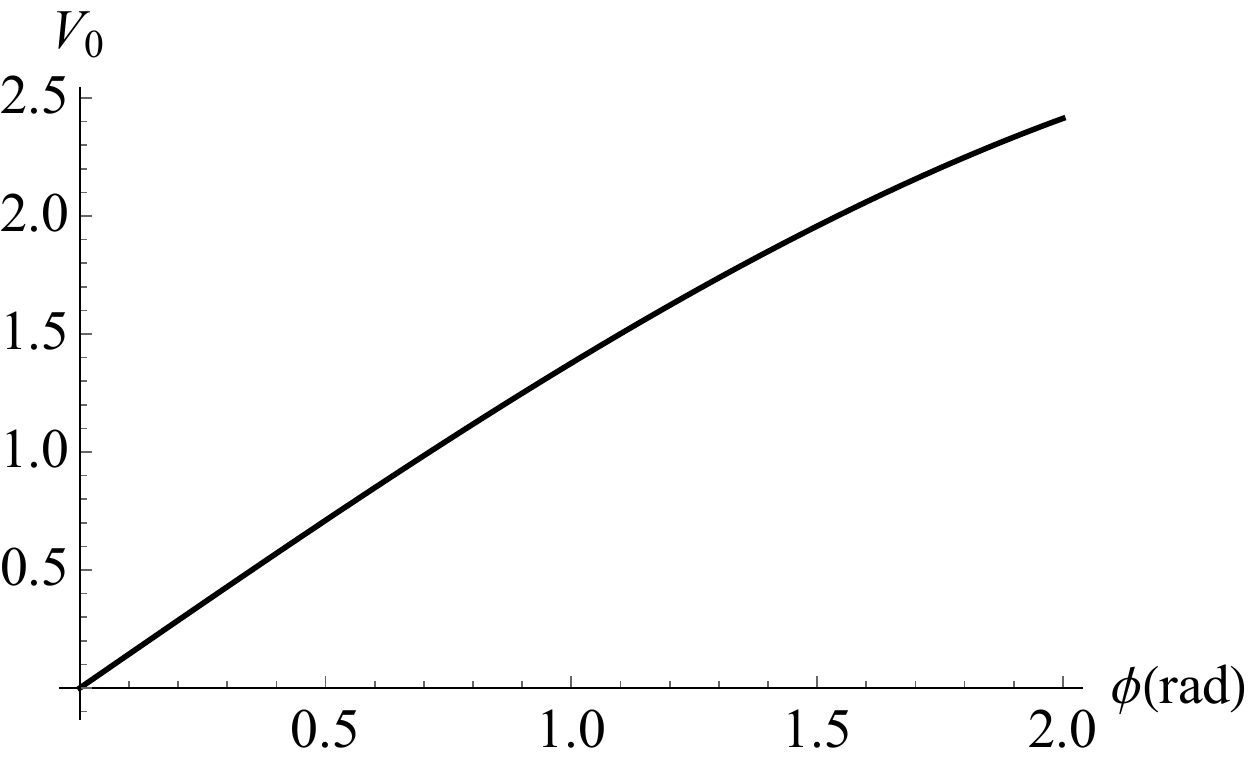}
          \includegraphics[width=0.5\textwidth]{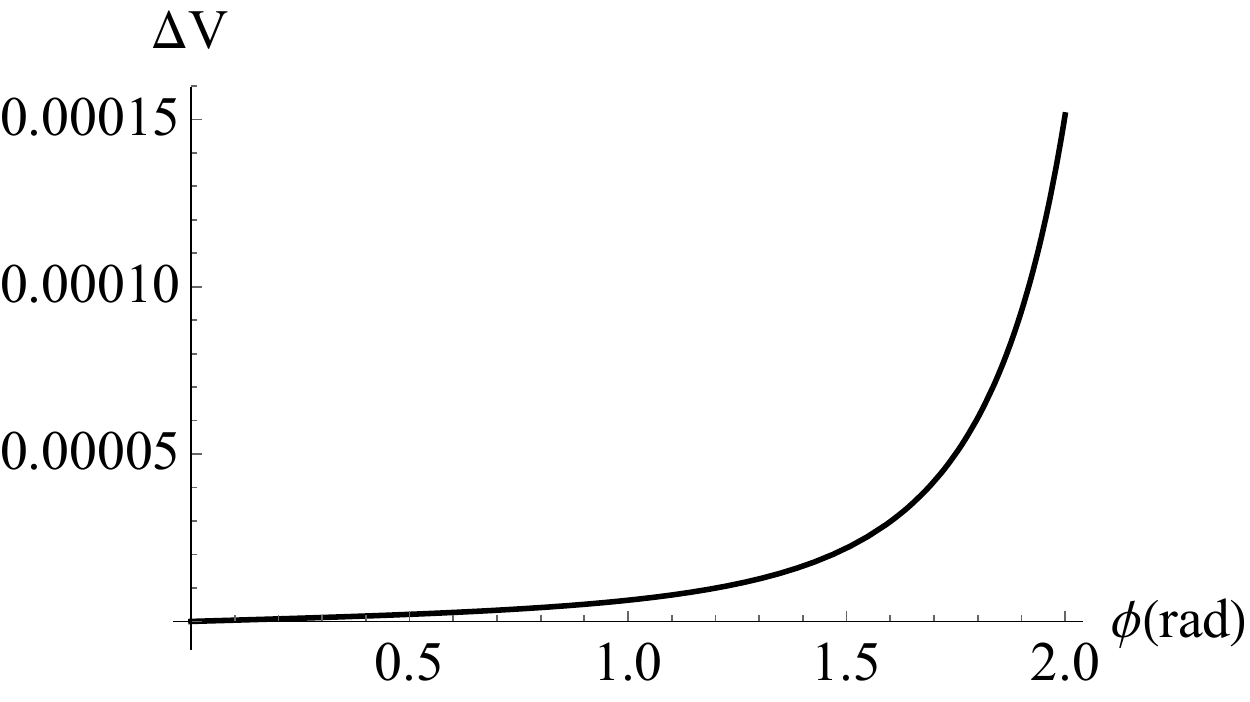}
      \caption{Left: the uncorrected potential $V_0$ as a function of $\phi$. The string worldsheet shrinks as $\phi \to 0$, so we have $V_0(\phi=0)=0$. Right: The $1/N$ energy correction \eqref{diffV} for $(N,\tilde{T})=(456,0.02)$.}
      \label{fig_diffV}
    \end{figure}

As an example, we choose $(N,\tilde{T})=(456,0.02)$, and show our result $\Delta V_{N,\tilde{T}}$ in Fig.~\ref{fig_diffV}.
For any value of $\phi$ the energy correction $\Delta V$ is found to be positive, thus the gravitational attraction is
weakened by the quantum gravity correction.
Therefore, we conclude that
{\it the $1/N$ correction provides a repulsive force}.
For some other values of $(N,\tilde{T})$ we evaluate the potential energy corrections and found again that
they are positive, so, as far as we search, the result is robust. 

In a sense it is reasonable to expect
that the quantum gravity correction may work as a screening of the gravitational force, thus weakening
the total gravitational effect.
An interesting note is that in Ref.~\cite{Hyakutake:2013vwa} the quantum gravity correction to the D0-brane potential 
due to the corrected spacetime \eqref{hyakutakemetric} was
found to be positive. This means that both the D0-brane and the fundamental string feel a repulsive force by the
quantum gravity correction of the spacetime.

\begin{figure}[t]
  		\centering
  		\includegraphics[width=0.5\linewidth]{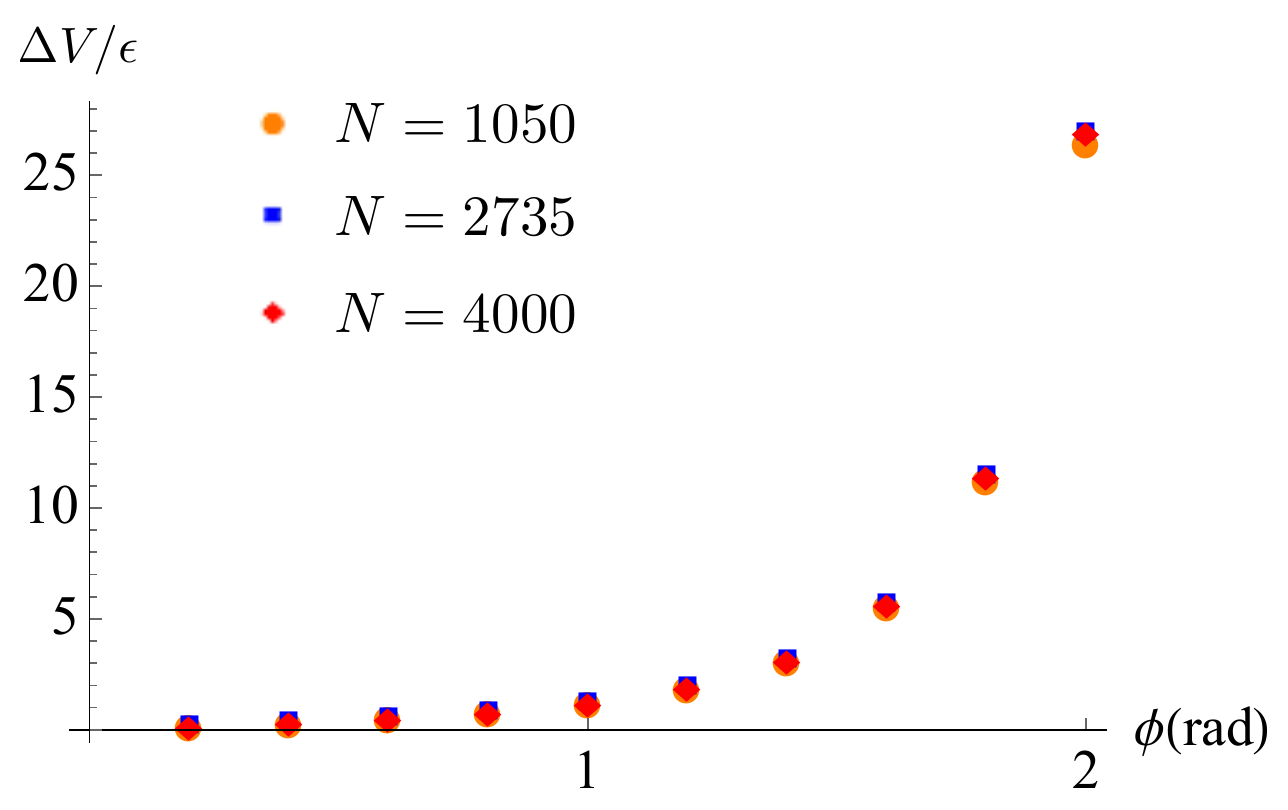}
  		\caption{Plot of $\Delta V_{N,\tilde{T}}/\epsilon$ for $N=1050,2735,4000$ with fixed $\tilde{T}=0.01$. They share the same $N^2$ dependence, as a check of our numerical calculations.
		}
  		\label{fig_Ndep}
  	\end{figure}

Let us finally check that the calculated correction $\Delta V$ behaves correctly as $1/N^2$. 
The quantum gravity correction of the metric \eqref{hyakutakemetric} is governed by the small parameter
$\epsilon \equiv \frac{\pi^6}{2^7 3^2} \frac{1}{N^2}$. So, the combination $\Delta V_{N,\tilde{T}}/\epsilon$ should not 
depend on $N$. In Fig.~\ref{fig_Ndep}, we plot $\Delta V_{N,\tilde{T}}/\epsilon$ for 
$N=1050,2735,4000$ with fixed $\tilde{T}=0.01$. As expected, the plots with different $N$ are found to be on top of each other.

\section{Summary}
\label{sec4}  
  
In this paper we calculated the $1/N$ corrections to the rectangular Wilson loop in the  
$SU(N)$ quantum mechanics with $16$ supercharges at finite temperature, by using the gauge/gravity correspondence.
The $1/N$ correction corresponds to the quantum gravity correction in the gravity side, so we introduced the 
quantum corrected 
near horizon geometry of the black 0-brane solution \eqref{hyakutakemetric} given in 
Ref.~\cite{Hyakutake:2013vwa}. We put a probe fundamental string
in the geometry, then the traditional holographic dictionary provided in Refs.~\cite{Maldacena:1998im,Rey:1998ik}
gave us the expectation value of the $1/N$ corrected Wilson loop. For the configuration of our probe string, see 
Fig.~\ref{fig_region} and Fig.~\ref{fig_image_string}.
  
For the holographic computations we needed to study the consistency conditions. First, we had to choose appropriate
temperature and $N$ as shown in Fig.~\ref{fig_NTregion} so that the quantum gravity corrections are dominant against
other corrections (see \eqref{N-Tregion1}) and also that the spacetime metric is real, \eqref{N-Tregion2}. Second, 
in some regions of the spacetime the background metric cannot be trusted as the correction overwhelmes the
classical geometry, so we need to work in a restricted spacetime region \eqref{xul}. 
The conditions for the region are given by \eqref{sugra_conditon1} and \eqref{sugra_conditon2}, and we adopt the 
threshold value $r=0.1$ to set the boundaries of the region. This determines the end points of the probe string,
which are on two probe D0-branes whose location corresponds to the vacuum expectation values of the scalar field
determined by the magnitude of the worldsheet instantons.

Our final result for the $1/N$ correction $\Delta V$ to the potential energy 
(which is the negative logarithm of the Wilson loop) is shown in Fig.~\ref{fig_diffV} Right. 
It turned out that the correction is positive for any angle $\phi$ between the two end points of 
the probe string. This means that the $1/N$ correction weakens
the potential. In the gravity side, the quantum gravity correction works as a repulsive force,
like a screening effect of the attractive gravitational force.

Finally, we make comments on future directions. The important check of holography concerning 
our results can be made by explicit calculations
in the gauge theory side. The D0-brane system admits numerical calculations of the observables 
\cite{Hanada:2007ti,Anagnostopoulos:2007fw,Hanada:2008ez,Hanada:2008gy}, so
we hope to see how the quantum gravity effect may be seen in gauge theories. 
Due to the validity region of the spacetime, we had to put two probe D0-branes
in the valid region so that the Euclidean string worldsheet lies safely for the calculation.
The precise location of the probe D0-branes needs to be specified by the string instanton amplitudes,
which would be a challenge for the precise numerical matching check of the AdS/CFT correspondence
for our case.

The understanding of the quantum gravity corrections is quite essential in holography. For example,
the correction to the entanglement entropy of the boundary theory is given by bulk spacetime entanglement
\cite{Faulkner:2013ana}. The corrections we considered in this paper has also been discussed in 
Ref.~\cite{Faulkner:2013ana} as a sub-leading correction. A unified understanding of the quantum gravity corrections
is necessary for a further investigation of the quantum gravity in the AdS/CFT correspondence, 
and our finding of the gravitational screening may serve as a hint for it.

\appendix

\section{Quantum gravity corrections to the near horizon geometry}

Here we provide an explicit expression of the quantum-corrected geometry \eqref{hyakutakemetric}
obtained in  Ref.~\cite{Hyakutake:2013vwa}.
The functions $H_i, F_1$ used to define the geometry are given as 
  \begin{align}
    H_{i} &\equiv \frac{(2\pi)^{4}15\pi\lambda^{-4/3}}{\tilde{U}_{0}^{7}}\left(
    \frac{1}{x^{7}}+\frac{\epsilon}{\tilde{U}_{0}^{6}}h_{i}
    \right),  \label{Hi} \\
    F_{1}&\equiv1-\frac{1}{x^{7}} +\frac{\epsilon}{\tilde{U}_{0}^{6}}f_{1}, \label{F1}
  \end{align}
  with the dimensionless $\tilde{U}_0\equiv U_0/\lambda^{1/3}$.
  The corrections are ${\cal O}(\epsilon)$ where $\epsilon \equiv \frac{\pi^6}{2^7 3^2} \frac{1}{N^2}$.
The functions appearing in these expressions are defined as follows.  
  \begin{align}
    h_{1}(x) &\equiv
    \frac{1302501760}{9x^{34}}-\frac{57462469}{x^{27}}+\frac{12051648}{13x^{20}}
    -\frac{482400}{12x^{13}} \nonumber\\
    &\qq{} -\frac{3747840}{x^{7}}+\frac{4099200}{x^{6}}-\frac{1639680(x-1)}{x^{7}-1}+117120\left( 18-\frac{23}{x^{7}}\right)I(x), \\
    h_{2}(x) &\equiv
    \frac{19160960}{x^{34}}-\frac{58528288}{x^{27}}+\frac{2213568}{13x^{20}}
    -\frac{1229760}{13x^{13}}\nonumber\\
    &\qq{} -\frac{2108160}{x^{7}}+\frac{2459520}{x^{6}}+1054080\left( 2-\frac{1}{x^{7}}\right)
    I(x), \\
    h_{3}(x) &\equiv \frac{361110400}{9x^{34}}-\frac{5840032}{x^{27}}-\frac{24021312}{13x^{20}}
    -\frac{58072000}{13x^{13}}\nonumber\\
    &\qq{} -\frac{210860}{x^{7}}+\frac{2459520}{x^{6}}+117120\left(
    1-\frac{41}{x^{7}}\right) I(x), \\
    f_{1}(x) &\equiv -\frac{1208170880}{9x^{34}}+　\frac{161405664}{x^{27}}+　\frac{5738880}{13x^{20}}+　\frac{956480}{x^{13}}+　\frac{819840}{x^{7}}I(x).
  \end{align}
The common function $I(x)$ is defined by 
  \begin{align}
    I(x)&\equiv \log{\frac{x^{7}(x-1)}{x^{7}-1}}-\sum_{n=1,3,5}\cos{\frac{n\pi}{7}}\log
    {\left( x^{2}+2x\cos{\frac{n\pi}{7}}+1\right) }\nonumber\\
    &\qq{} -2\sum_{n=1,3,5}\sin{\frac{n\pi}{7}}\left\lbrace \tan^{-1}\left(
    \frac{x+\cos{\frac{n\pi}{7}}}{\sin{\frac{n\pi}{7}}}\right) -\frac{\pi}{2}
    \right\rbrace.
  \end{align}

\begin{figure}[t]
  \centering
    \includegraphics[width=.49\linewidth]{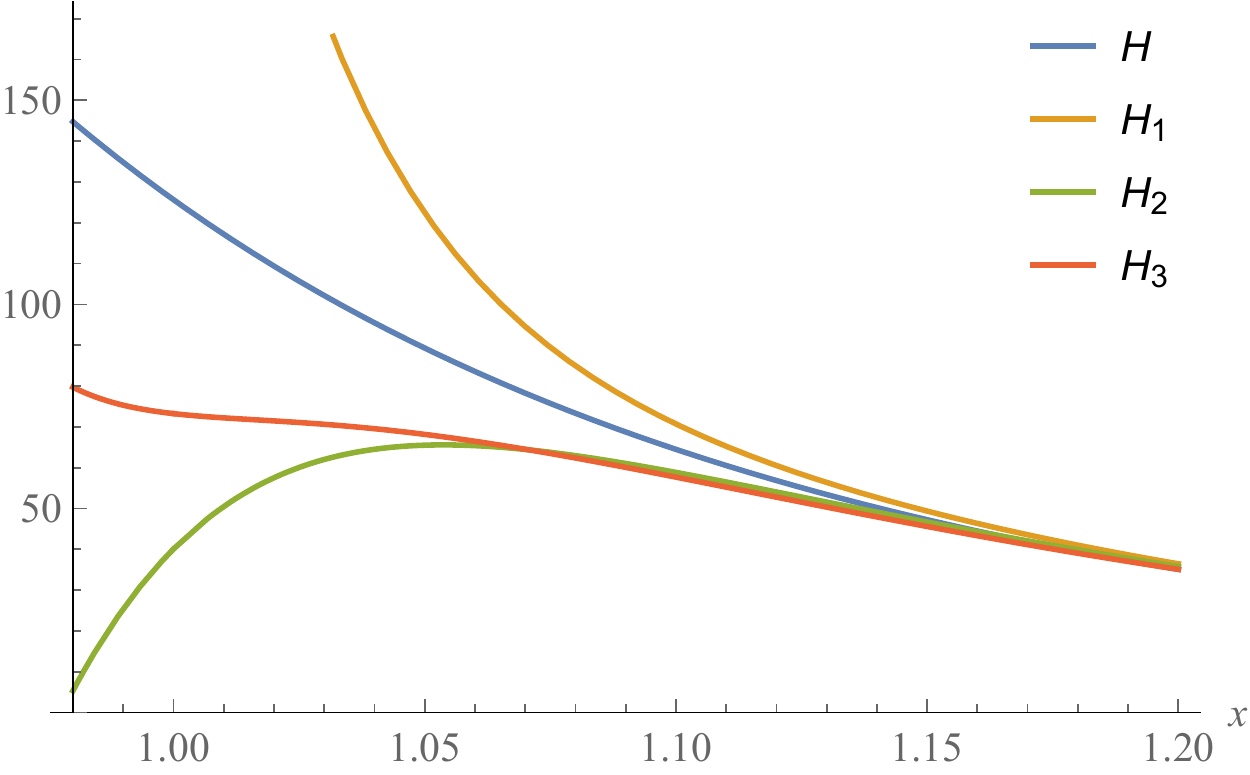}
    \includegraphics[width=.49\linewidth]{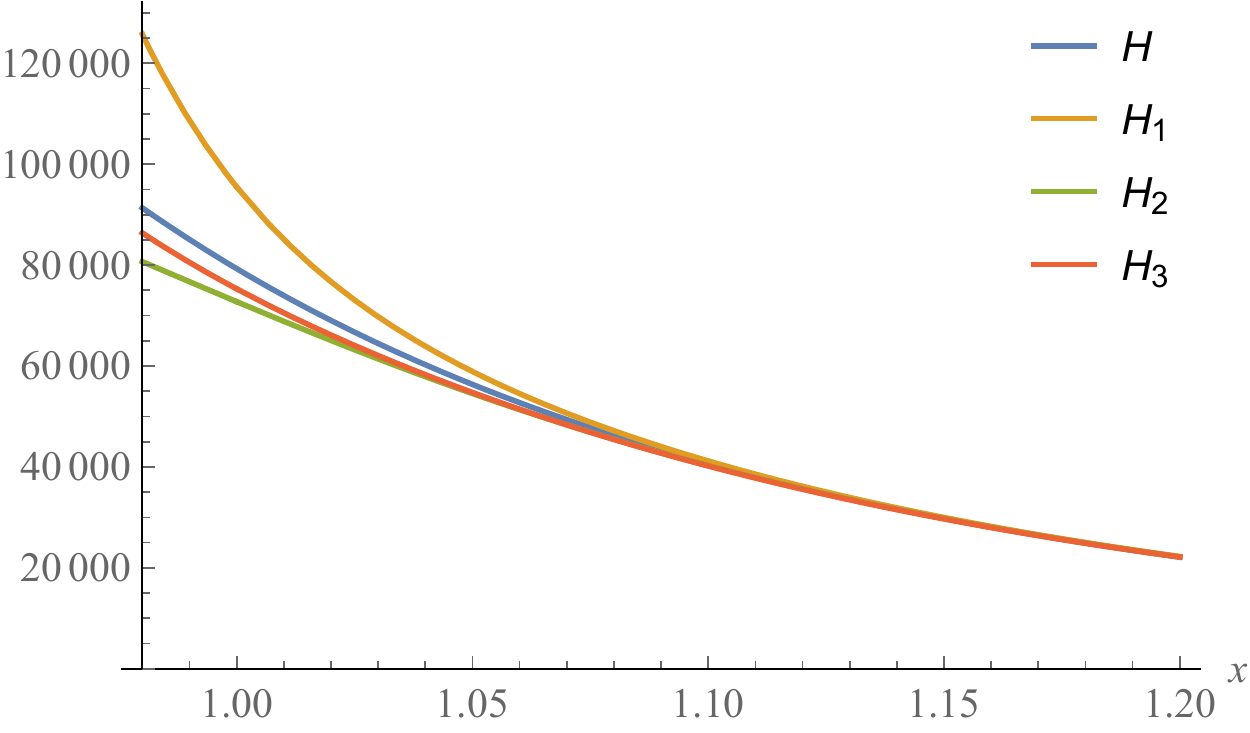}
  \caption{The behavior of the functions in the metric components. $H(x)$ is the classical geometry without the quantum correction, 
  and $H_1(x), H_2(x)$ and $H_3(x)$ are with the corrections. Left:  The case with $N  = 456 $ and  $ \tilde{T} =0.02.$ Right: 
  The case with $N=20000$ and $ \tilde{T} =0.002$.
  }\label{fig_HandHi}
  \end{figure}

The functions $H_i$ and $F_1$ approach $H$ and $F$ in the limit $N\to \infty$ or at large  $x$. 
Near $x\sim 1$, the correction terms are effective. See Fig.~\ref{fig_HandHi} for the behavior of the functions with the
quantum corrections.

\acknowledgments

We would like to thank T.~Akutagawa, Y.~Hyakutake and 
N.~Iizuka for valuable discussions.
The work of K.H. is supported in part by JSPS KAKENHI Grants No. JP15H03658, No. JP15K13483, and No. JP17H06462.

\bibliographystyle{JHEP}
\bibliography{paper}

%
%
%
%
%
%
%
%
%

\end{document}